\documentclass[twocolumn,showpacs,preprintnumbers,amsmath,amssymb]{revtex4}

\usepackage{graphicx}
\usepackage{dcolumn}
\usepackage{bm}

\begin{document}


\title{Anomalous impurity effect on magnetization in frustrated 
one-dimensional ferro- and ferrimagnets}

\author{Masanori Kohno and Xiao Hu}
\affiliation{Computational Materials Science Center, National Institute 
for Materials Science, Tsukuba 305-0047, Japan}

\date{\today}

\begin{abstract}
Significant decrease of spontaneous magnetization in frustrated 
one-dimensional ferro- and ferrimagnets due to non-magnetic impurities is 
predicted. Using the density-matrix renormalization group method and the exact 
diagonalization method, we confirm that the total spin can vanish due to 
a single impurity in finite chains. Introducing the picture of magnetic 
domain inversion, we numerically investigate the impurity-density 
dependence of magnetization. In particular, we show that even with an 
infinitesimal density of impurities the magnetization in the ground state 
is reduced by about 40\% from that of the corresponding pure system. 
Conditions for the materials which may show this anomalous impurity effect 
are formulated.
\end{abstract}


\maketitle

Frustrations in quantum spin systems have attracted much attention for its 
potential to exhibit new phenomena which have never been observed in 
unfrustrated systems. Various possibilities in frustrated systems have been 
suggested, such as exotic excitations near critical points\cite{Senthil}, 
incommensurate orders in magnetic fields\cite{eta_inv}, chiral 
orderings\cite{chiral_order_Exp,chiral_order_Num1,chiral_order_Num2,
chiral_order_Num3} and disordered ground states\cite{RVB1,RVB2}. 
The property we will discuss in this paper is also one of the phenomena 
where frustrations play an essential role, and will never be observed 
in unfrustrated systems. The property is an impurity effect on frustrated 
one-dimensional ferro- and ferrimagnets.
\par
Usually, a small amount of impurities has little influence on bulk magnetic 
quantities, since the mean distance between impurities is so long that the 
correlation between them is very weak and usually they affect only local 
quantities. However, in some special situations, a small amount of impurities 
can cause a bulk effect. An example is the impurity-induced antiferromagnetic 
long-range order (AFLRO) in quasi-one-dimensional spin-1/2 spin-gap systems, 
which was thoroughly investigated by 
theoretical\cite{SGimp_theo1,SGimp_theo2,SGimp_theo3}, 
numerical\cite{SGimp_num1,SGimp_num2} and 
experimental\cite{SGimp_exp1,SGimp_exp2} approaches. 
The main feature of this effect is roughly explained as follows: 
Without an impurity, spins form dimers locally. By introducing 
non-magnetic impurities, moments are induced around impurity sites. 
The moments couple one to another and exhibit antiferromagnetic alignment. 
Hence, AFLRO and low-energy spin-wave excitation appear in the background 
of high-energy triplet excitation. For this effect, the spin gap and the 
correlations between induced moments play an essential role. In this paper, 
we present another example that a bulk quantity, magnetization, is 
substantially influenced by a small amount of impurities due to a different 
mechanism from that of the impurity-induced AFLRO. The impurity effect will 
be realized even with an infinitesimal density of impurities in frustrated 
ferro- and ferrimagnetic chains that satisfy the conditions we will present 
in this paper. Thus, in such systems the magnetization in the ground state 
will be significantly reduced from that of the corresponding pure systems 
even with a usually-negligible amount of impurities. We notice that without 
the knowledge developed in this work, reduction of magnetization from the 
expected values tends to be explained by assuming complex, higher order 
interactions such as the Dzyaloshinsky-Moriya interaction or interchain 
antiferromagnetic couplings. 
\par
\begin{figure}
\includegraphics[scale=0.15]{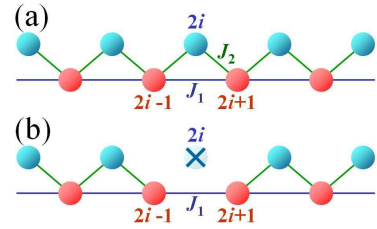}
\caption{Lattice structure of models 1 and 2 (a) without an impurity and 
(b) with an impurity at site $2i$.}
\label{fig:lattice}
\end{figure}

In order to demonstrate the anomalous effect of non-magnetic impurities on 
frustrated ferro- and ferrimagnets, we consider the minimal models defined 
by the following Hamiltonian:
\begin{equation}
{\cal H}=J_1\sum_i{\mbox {\boldmath $S$}}_{2i-1}\cdot
{\mbox {\boldmath $S$}}_{2i+1}+J_2\sum_i{\mbox {\boldmath $S$}}_{i}\cdot
{\mbox {\boldmath $S$}}_{i+1},
\end{equation}
where ${\mbox {\boldmath $S$}}_i$ denotes the spin operator at site $i$. 
The lattice structure is shown in Fig. \ref{fig:lattice} (a).
In model 1, the spin lengths of all spins are one half. The coupling constant 
$J_1$ is antiferromagnetic ($J_1$$>$0), and $J_2$ is 
ferromagnetic ($J_2$$<$0). 
In model 2, the spin lengths at even sites are one half, and those at odd 
sites are one; both coupling constants are antiferromagnetic ($J_1$ and 
$J_2$$>$0). Model 1 is nothing but the one proposed by Hamada and his 
coworkers in Ref. \cite{Hamada}. Model 2 can be reduced to well-known models 
by neglecting $J_1$ or $J_2$: At $J_2$=0 this model is equivalent to the $S$=1 
Heisenberg chain and free spins, and at $J_1$=0 it is nothing but the 
spin-alternating Heisenberg chain. The ground states of models 1 and 2 become 
ferro- and ferrimagnetic, respectively, when $|J_2|$ is sufficiently larger 
than $J_1$. Hereafter, the number of unit cells and the number of sites are 
denoted by $L$ and $N_s$, respectively, and open boundary conditions are 
applied. 
In models 1 and 2, $L$=$N_s$/2.
\par
\begin{figure}
\includegraphics[scale=0.19]{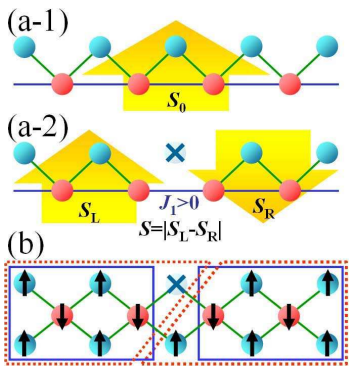}
\caption{(a) Schematic picture of magnetic domains (a-1) without an impurity 
and (a-2) with an impurity. (b) Ferrimagnetic chain on a bipartite lattice, 
where the spin lengths of all spins are one half, and all coupling constants 
are antiferromagnetic.}
\label{fig:domain}
\end{figure}

When a non-magnetic impurity is doped at an odd site, the system is 
decomposed into two pure systems. Then, the situation is rather trivial. 
Thus, we first concentrate on the case 
where an impurity occupies an even site as shown in Fig. \ref{fig:lattice} 
(b). Since the interactions from the impurity site are removed, the remaining 
interaction between the spins adjacent to the impurity site is $J_1$, 
which is antiferromagnetic. Hence, we expect that the total spin $S$ in the 
ground state, which corresponds to the spontaneous magnetization, becomes 
$|S_L$$-$$S_R|$, where $S_L$ and $S_R$ are those of the blocks to the 
left and right of the impurity site, respectively. The expected picture of 
this inversion of magnetic domains is schematically shown in 
Fig. \ref{fig:domain} (a). As a special case where $S_L$=$S_R$, 
the total spin will vanish due to a single impurity. 
\par
In order to confirm this substantial decrease in spontaneous magnetization, 
we calculated $S$ in the ground states of models 1 and 2 with a single 
impurity put at all possible even sites in up to 40-site chains with even $L$ 
by the density-matrix renormalization group (DMRG) method\cite{DMRG} and the 
exact diagonalization method. The coupling constants are set to be 
$J_1$=0.1 and $|J_2|$=1. We calculated the total spin $S$ in the ground state 
by using the formula 
$S$($S$+1)=$\langle\left(\sum_{i}{\mbox {\boldmath $S$}}_i\right)^2\rangle$
(=$\langle\sum_{i,j}{\mbox {\boldmath $S$}}_i$$\cdot$${\mbox {\boldmath $S$}}_j
\rangle$), where $i$ and $j$ run over all sites, and $\langle$ $\rangle$ 
denotes the expectation value in the ground state. The numerical results on 
the total spin $S$ satisfied the relation $S$=$|S_L$$-$$S_R|$ in all 
the cases we have investigated. There is 
no mathematical proof on this relation for quantum spin systems with 
frustrations, hence it is nontrivial. The physical picture of the 
domain inversion can be intuitively understood by considering the 
corresponding Ising models, for which this relation holds with total spin 
$S$ replaced with total $z$-component of spins $S^z$.
\par
It should be noted, on the other hand, that in unfrustrated ferrimagnetic 
chains on bipartite lattices such as shown in Fig. \ref{fig:domain} (b), this 
anomalous impurity effect does not occur, since the sign of the effective 
coupling between domains does not change due to impurities: In both 
definitions of domains denoted by solid and dotted lines in 
Fig. \ref{fig:domain} (b), the sign of the effective coupling 
remains the same before or after impurity doping. Actually, in these systems, 
ferrimagnetic ground states are ensured by the Marshall-Lieb-Mattis 
theorem\cite{Marshall,Lieb_Mattis} with or without an impurity. 
\par
\begin{figure}
\includegraphics[scale=0.31]{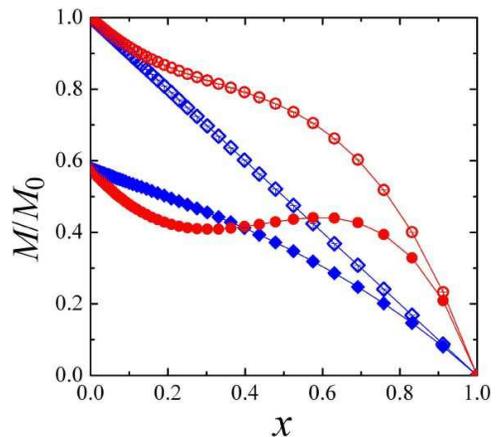}
\caption{Impurity-density dependence of magnetization. The magnetizations 
without an impurity are denoted by $M_0$. Solid diamonds and circles denote 
the results on models 1 and 2, respectively. Open symbols denote those of the 
corresponding unfrustrated systems by setting $J_1$ to be ferromagnetic.}
\label{fig:xdep}
\end{figure}

Based on the above numerical results for models 1 and 2 doped with an 
impurity, it is natural to expect that the total spin $S$ is expressed 
in terms of those in domains ($S_k$) as 
\begin{equation}
S=\left|\sum_k (-1)^k S_k\right|, 
\end{equation}
when impurities are doped at even sites. Taking this relation into account, 
we have calculated magnetization $M$ in an infinitesimal magnetic field with 
impurities randomly distributed on a chain, where impurities can sit not only 
on even sites but also on odd sites. Here, the magnetization $M$ in an 
infinitesimal magnetic field is expressed in terms of the total spin $S_l$ in 
the $l$-th isolated cluster as $M$=$\sum_lS_l$. To be concrete, we have 
calculated the average of $M$ over 10,000 chains. Each chain has 100/$x$ 
sites and 100 randomly distributed impurities, where $x$ is the impurity 
density. The numerical result on the impurity-density dependence of 
magnetization is shown in Fig. \ref{fig:xdep}. 
Magnetizations of models 1 and 2 are drastically reduced due to impurities 
(solid diamonds and circles, respectively). In particular, 
in the limit of small impurity-density, the magnetizations decrease down to 
about 57.7\% of those of the corresponding pure systems; 
$M$($x$$\rightarrow$0)$\simeq$0.577$\times$$M$($x$=0). 
\par
This feature is contrasted with that without frustrations: As an example, 
we consider the model 1 with all coupling constants ferromagnetic. In this 
model, the ground state is ferromagnetic with or without an impurity. 
Thus, the magnetization decreases by the amount of the spins at 
impurity sites. Namely, the magnetization linearly decreases as a function of 
the impurity density, i.e. $M/M_0$=1$-$$x$, as shown in Fig. \ref{fig:xdep} 
(open diamonds). Since the corresponding Ising models show the same impurity 
effect for the $z$-component of spins, it is expected that the effect shown 
here will also be realized in XXZ models with Ising-like anisotropy. 
\par
\begin{figure}
\includegraphics[scale=0.12]{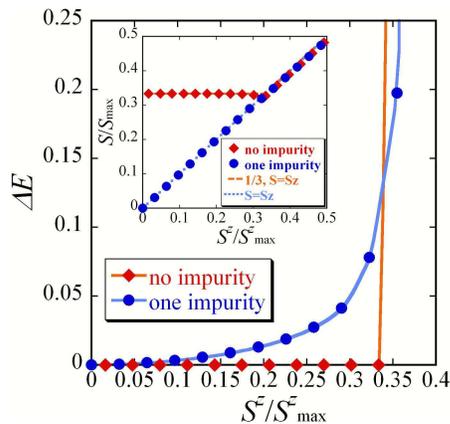}
\caption{Ground-state energies as a function of $S^z$ measured from that of 
$S^z$=0 in a 42-site cluster of model 2 with $J_1$=0.1 and $J_2$=1 without 
an impurity (diamonds) and with an impurity at the 20-th site (circles).
Solid lines are guide to eyes. The inset shows the calculated $S$. 
Dashed and dotted lines indicate typical behaviors of a ferrimagnet and 
a disordered state, respectively. The $S$ and $S^z$ of the fully polarized 
state are denoted by $S_{\rm max}$ and $S^z_{\rm max}$, respectively.}
\label{fig:EM}
\end{figure}

Now, let us consider excitations in doped systems with non-magnetic 
impurities. We calculated energies of a 42-site chain of model 2 with open 
boundary conditions, where an impurity is put at the 20-th site. The coupling 
constants are set to be $J_1$=0.1 and $J_2$=1. The finite-size algorithm of 
the DMRG method\cite{DMRG} is applied with truncation number up to $m$=$150$. 
We performed 10 sweeps and confirmed convergence by calculating 
$S$($S$+1)=$\langle\sum_{i,j}{\mbox {\boldmath $S$}}_i$$\cdot$$
{\mbox{\boldmath $S$}}_j\rangle$. In the inset of Fig. \ref{fig:EM}, we plot 
the calculated $S$ with respect to total $z$-component of spins 
$S^z$(=$\sum_iS_i^z$). The calculated values of $S$ almost coincide with the 
typical behaviors of a ferrimagnet and a disordered state (dashed and dotted 
lines, respectively), indicating that the wavefunctions with various $S^z$'s 
are well converged. This figure also shows that the ground state changes from 
a ferrimagnetic state to a spin-singlet state due to a single impurity. 
The ground-state energies within the subspaces of fixed $S^z$ measured from 
that of $S^z$=0 are shown in Fig. \ref{fig:EM}. 
\begin{figure}
\includegraphics[scale=0.12]{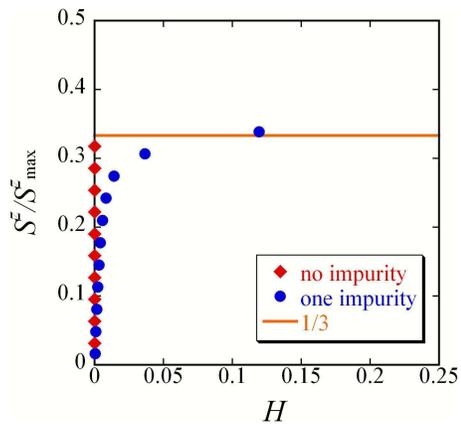}
\caption{Magnetization curve of model 2. The parameters are the same as those 
in Fig. \ref{fig:EM}.}
\label{fig:MH}
\end{figure}
This figure suggests that there is a low-energy continuous excitation 
from a spin-singlet state\cite{size_gap}. This feature is contrasted with 
that of the Ising model where the lowest excitation has a finite gap of the 
order of $J_1$ or $J_2$ independently of the cluster size. The ground states 
with small $S^z$'s in doped quantum systems are almost degenerate, 
which would be reflecting ferromagnetic fluctuations in domains.
\par
We calculated magnetic field $H$ by using a discretized form of the 
derivative of energy $E$ with respect to $S^z$: $H$=
$\partial E/\partial S^z$$\simeq$$\{E(S^z_{n+1})$$-$$E(S^z_n)\}
/\{S^z_{n+1}$$-$$S^z_n\}$, where $S^z_n$=$n$ or $n$$+$0.5 with or without an 
impurity. ($n$=0, 1, $\cdots$.) 
The result on the magnetization curve is 
shown in Fig. \ref{fig:MH}. The magnetic field required for the magnetization 
to recover up to the spontaneous magnetization of the pure system is about 
0.1 which is the order of $J_1$ as expected from the picture of domain 
inversion (Fig. \ref{fig:domain} (a)). 
\par
Based on the above considerations, we list the conditions for the anomalous 
impurity effect:
\begin{enumerate}
\item The system should be one-dimensional. Namely, interactions between 
chains should be much smaller than those in chains.
\item The ground state without an impurity should have spontaneous 
magnetization.
\item Local interactions near impurity sites should be set such that the 
effective interaction between magnetic domains changes from ferromagnetic to 
antiferromagnetic due to impurities. 
\end{enumerate}
The third condition leads to frustration. 
\par
The models that satisfy the above conditions will exhibit the anomalous 
impurity effect. 
\begin{figure}
\includegraphics[scale=0.228]{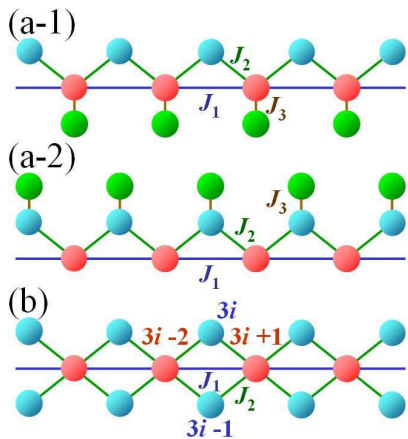}
\caption{Possible models for the anomalous impurity effect with spins 
one half and coupling constants antiferromagnetic. 
(a) Decorated triangle chains. (b) Diamond-like chain.}
\label{fig:S05mdl}
\end{figure}
For example, the decorated triangle chains (Figs. \ref{fig:S05mdl} 
(a-1) and (a-2)) and the diamond-like chain (Fig. \ref{fig:S05mdl} (b)) 
will be the models that exhibit this effect with all spins one half and all 
coupling constants antiferromagnetic. In the decorated triangle chains, 
when $J_2$ is sufficiently larger than $J_1$, the spins on decorating sites 
align parallel, resulting in a ferrimagnetic ground state. If an impurity is 
doped on the top site of a triangle, the remaining interaction between the 
spins adjacent to the impurity is $J_1$, which is antiferromagnetic. Thus, 
the domain inversion and substantial decrease in magnetization due to 
impurities are expected. Actually, we have confirmed by exact diagonalization 
that the total spin $S$ in the ground state behaves as 
$S$=$|S_L$$-$$S_R|$, when an impurity is doped on top sites of 
triangles in up to 24-site clusters with $J_1$=0.1, $J_2$=1.0 and 
$J_3$=0.5\cite{model_a2}. 
\par
In the case of the model in Fig. \ref{fig:S05mdl} (b), the ground state 
becomes ferrimagnetic, when $J_2$ is sufficiently larger than $J_1$. 
If an impurity is doped at site 3$i$, the effective interaction between the 
spins at sites 3$i$$-$2 and 3$i$$+$1 is mainly determined by the three-site 
Hamiltonian of sites 3$i$$-$2, 3$i$$-$1 and 3$i$$+$1. In order for the 
effective interaction to be antiferromagnetic, $J_1$ has to be larger than the 
effective coupling by $J_2$'s through the spin at site 3$i$$-$1. If such a 
parameter can be chosen, the magnetic domain inversion due to impurities 
will be realized. In this paper, we do not intend to determine the precise 
boundaries for this effect, since in delicate systems such as that of 
Fig. \ref{fig:S05mdl} (b) the phase boundary will depend on the system size. 
Instead, we would like to emphasize that, as demonstrated in this paper, 
there actually exist systems that exhibit this impurity effect in some 
parameter regimes for frustrated ferro- and ferrimagnets in one dimension.
\par
In summary, we have investigated effects of non-magnetic impurities on 
frustrated ferro- and ferrimagnets in one dimension by the DMRG method and 
the exact diagonalization method. Based on the numerical results, we pointed 
out that in these systems a small amount of impurities can drastically 
decrease magnetization in the ground state. Introducing the picture of 
magnetic domain inversion, we have investigated impurity-density dependence of 
magnetization. In particular, we have shown that the magnetization with an 
infinitesimal density of impurities becomes as small as 57.7\% of that 
without an impurity. The energy scale of this impurity effect is of the order 
of the remaining effective interaction between the spins adjacent to 
impurity sites. The low-energy excitations in doped systems are continuous 
from the lowest spin-state (except the finite-size gap). We also listed 
the conditions for this impurity effect. 
\par
In the materials which are effectively described by frustrated spin models, 
other interactions such as the Dzyaloshinsky-Moriya interaction or 
biquadratic interactions are sometimes not negligible. Although their 
influence on the impurity effect requires further study, the prediction 
in this paper deserves careful experimental investigations. 
\par
We thank M. Hase for valuable discussions and comments on related materials.

\bibliography{apssamp}

\begin{references}
\bibitem{size_gap}In finite-size systems, there should be a gap due to the 
discreteness of energy levels. Here, we mean that an asymptotically 
continuous low-energy excitation exists which reflects the quantum nature 
of the system.
\bibitem{model_a2}For the model in Fig. \ref{fig:S05mdl} (a-2), the degrees 
of freedom of the spin on the decorating site connected to the impurity are 
neglected, since it behaves as a free spin.
\end{references}

\end{document}